\documentstyle[aps,pre,epsfig,graphicx]{revtex}
\newcommand{\be}{\begin{equation}}
\newcommand{\bea}{\begin{eqnarray} \nonumber}
\newcommand{\ee}{\end{equation}}
\newcommand{\eea}{\end{eqnarray}}

 \def\(({\left(}
 \def\)){\right)}
\def\[[{\left[}
\def\]]{\right]}
\def\bi{\bibitem}
\def \form#1 {eq. (\ref{#1}) }
\def \parziale#1#2  {{\partial {#1} \over \partial {#2}}}

\def \cN{{\cal N}}

\def \ba#1 {\overline{#1}}

\def  \s {\sigma}
\def \vp {{\vec p}}
\def \cG {{\cal G}}
\def \cQ {{\cal Q}}
\def \cP {{\cal P}}
\def \eps{\epsilon}

\begin{document}

\title{The cavity method at zero temperature}

\author{Marc M\'ezard}
\address{Laboratoire de Physique Th\'eorique et Mod\`eles Statistiques
\\
Universit\'e Paris Sud, Bat. 100, 91405 Orsay {\sc cedex}, France \\
mezard@ipno.in2p3.fr}
\author{Giorgio Parisi}
\address{Dipartimento di Fisica, Sezione INFN, SMC and UdRm1 of INFM,\\
Universit\`a di Roma ``La Sapienza'',
Piazzale Aldo Moro 2,
I-00185 Rome (Italy)\\
giorgio.parisi@roma1.infn.it}

\date{\today}

\maketitle
\begin{abstract}
In this note we explain the use of the cavity method directly at zero
temperature, in the case of the spin glass on a
lattice with a local tree like structure, which is the proper generalization
of the usual Bethe lattice to frustrated problems. 
The computation is done
explicitly in the formalism equivalent to `one step replica symmetry
breaking'; we compute the energy of the global ground state, as well as the
complexity of equilibrium states at a given energy. Full results are presented
for a Bethe lattice with connectivity equal to three.
The main assumptions underlying the one step cavity approach, namely the existence 
of many local ground states, are explicitely stated and discussed:
some of the main obstacles towards a rigorous study of the
problem with the cavity method are outlined.

\end{abstract}

\section{Introduction}
The cavity method, initially invented to deal with the Sherrington Kirkpatrick
model of spin glasses \cite{MPV_cav,MPV}, is a powerful method to compute the
properties of ground states in many condensed matter and optimization
problems. It was originally developed at finite temperature, but if one is
mainly interested in the optimization problem concerning the structure of
ground states, it turns out to be easier to apply it directly at zero
temperature, where many of the concepts can be explained in a more
straightforward way. The aim of this note is to discuss the use of the cavity
method at zero temperature. In order to be definite, we will study the
concrete case of a spin glass on the Bethe lattice 
(the precise definition of the Bethe lattice which we use for spin glasses is
given in sect. II),
 although the scope of
application of the method is much wider; indeed the best applications of the
formalism we present here are those dealing with optimization problems
\cite{optim_revue}, like for instance the K-satisfiability problem \cite{Ksat}
or some of its variants \cite{pspin}.

The cavity method is in principle equivalent to the replica method, but it
turns out to have a much clearer and more direct interpretation, that allows
in practice to find solutions to some problems which remain rather difficult
to understand in the replica formalism: the replica approach is very elegant
and compact, but it is more difficult to get an intuitive feeling of what is
going on. Also, the cavity approach deals with usual probabilistic objects,
and can lend itself to rigorous studies \cite{talag,bovier}. We shall present
it here at two successive levels of approximation. The first one,
corresponding in replica language to the replica symmetric (RS) solution, is an
easy one and has already been studied a lot. The main aim of the paper is to
explain in some details how one can solve the problems at the level of a 'one
step replica symmetry breaking' (1RSB). For years, this was only possible for
systems with infinite range interactions. We have recently found a general
procedure allowing to get this 1RSB solution for problems defined on the Bethe
lattice \cite{MP_Bethe}, and here we explain in some details its use directly
at zero temperature. We shall give an explicit example of this 1RSB solution
for the Bethe lattice spin glass with connectivity equal to three.

The paper is organised as follows: in sect. \ref{gener} we introduce the Bethe
lattice spin glass, the basic ideas of iterative techniques, and the main
questions which one wants to answer. Sect. \ref{RSsect} presents the cavity
method when one assumes that there is only one local ground state (the RS
approximation). The 1RSB solution is described in sect. \ref{1RSBsect}. The paper
ends with some remarks (sect. \ref{conclsect}), and is complemented by two
appendices: Appendix A shows how the present zero
temperature work is connected to the limit $T\to 0$ of finite temperature
analyses, and how this can be used to estimate the validity of a given level
of RSB. Appendix B explains how the cavity equations can be interpreted
from a variational point of view. Appendix C contains some miscellaneous
comments on the definition of the local ground states.

\section{Generalities}
\label{gener}
We consider an Ising spin glass model on a special class of random lattices
which we call Bethe lattice. This is defined here as as a random lattice with
fixed connectivity $k+1$. Such graphs locally look like a portion of a Cayley
tree, but they display loops at large distances (of length of order
$\log(N)/\log(k)$, where $N$ is the number of points in the lattice). In
unfrustrated problems, the Bethe lattice is defined as the interior of a large
Cayley tree, but this definition is not appropriate for frustrated systems
like spin glasses, where the presence of loops is essential to insure the
existence of frustration. It has been argued in \cite{MP_Bethe} that the
natural generalization of the Bethe lattice to frustrated systems is provided
by the random lattices with fixed connectivity which we study here.
 Notice
that in these systems, the frustration and the disorder are due to the
presence of loops, and thus occur only on large scales. Locally, the structure
around almost all point is tree-like.

Spins are located at the vertices of the
graph, and interact with neighbouring spins with exchange couplings.
In the spin glass case the Hamiltonian  is defined as:
\be
H=-\sum_{<ij>} J_{ij} \sigma_i \sigma_j \ .
\ee
The sum is over all links of the lattice (there are $k+1$ links incoming
onto each site $i$).  
For each link $<ij>$ the coupling $J_{ij}$ is an 
independent random variable chosen with the same probability distribution $P(J)$.

One of our aims is to compute, in the infinite $N$ limit, the 
value of the energy density of 
the global ground state (GGS), which is the configuration of Ising spins 
$\sigma_i=\pm 1$ which minimizes the Hamiltonian.
More precisely
the ground state energy of a $N$ spin system, averaged 
over the distribution of samples (i.e. both over the choices of random graphs,
and the values of the couplings) will be denoted by $E_N$. We want to compute
\be
U=\lim_{N\to\infty}{ E_{N} \over N} \ .
\ee

We are also interested in computing, in the same limit, the number of  local ground states
(LGS) with a given energy, where a LGS is defined as a configuration whose 
energy cannot be decreased by flipping a finite  number of spins 
(when $N \to \infty$).

\begin{figure}[bt]
\centerline{    \epsfysize=6cm
       \epsffile{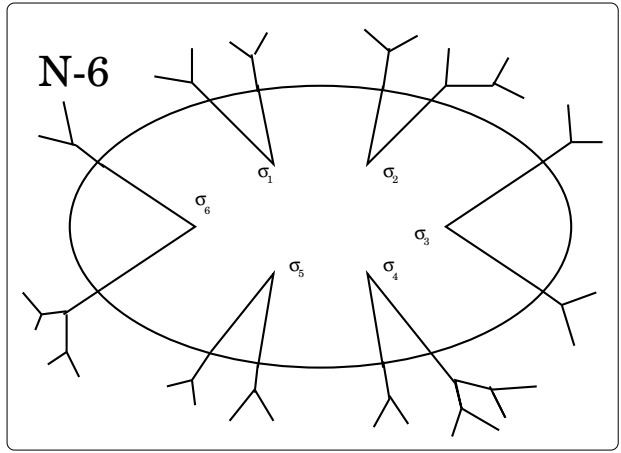}}
\caption{An example, for the case $k=2$, of a  $\cG_{N,6}$ cavity
graph where
$q=6$ randomly chosen cavity spins have  $2$ neighbours only. All the other 
$N-6$ spins  outside the cavity are connected through a random
graph such that every spin has $k+1=3$ neighbours.
\label{fignmoinssix}}
\end{figure}

The basic locally tree-like structure is best exploited by an iterative method that has been called, 
in the context of spin glasses, the cavity method.  Let us introduce an intermediate object which is 
a spin glass model with $N$ spins, on a slightly different random lattice, where $q$ randomly chosen 
'cavity' spins have only $k$ neighbours, while the other $N-q$ spins all have $k+1$ neighbours (see 
fig.\ref{fignmoinssix}).  We call such a graph a $\cG_{N,q}$ 'cavity graph'.  The cavity spins are 
fixed, their values are $\sigma_1,...,\sigma_{q}$.  The GGS energy of the corresponding spin glass model 
obviously depends on the values of the cavity spins.

While our primary interest is in the ground state configurations on $\cG_{N,0}$
graphs, the intermediate construction of  $\cG_{N,q}$ is helpful.
The basic operations which one can perform on cavity graphs are the following:
\begin{itemize}
\item Iteration:
By adding a new spin $\sigma_0$ of fixed value into the cavity,  connecting it
to   $k$
of the cavity spins say $\sigma_1,...,\sigma_k$, and optimizing  the
values of these $k$ spins,
one  changes a  $\cG_{N,q}$ into a  $\cG_{N+1,q-k+1}$ graph:
\be
\delta N =1, \ \ \ \ \ \  \delta q =-k+1 \ .
\ee
\item Link addition:
By adding a new link between two randomly chosen cavity spins $\sigma_1,\sigma_2$,
and optimizing  the
values of these $2$ spins
one changes a  $\cG_{N,q}$ into a  $\cG_{N,q-2}$ graph:
\be
\delta N =0, \ \ \ \ \ \  \delta q =-2 \ .
\ee
\item Site addition:
By adding a new spin $\sigma_0$ into the cavity, connecting it
to   $k+1$
of the cavity spins say $\sigma_1,...,\sigma_{k+1}$,
 and optimizing  the
values of the $k+2$ spins $\sigma_1,...,\sigma_{k+2}$,
one  changes a  $\cG_{N,q}$ into a  $\cG_{N+1,q-k-1}$ graph:
\be
\delta N =1, \ \ \ \ \ \  \delta q =-k -1 \ .
\ee
\end{itemize}
In particular, if one starts from a $\cG_{N,2(k+1)}$ cavity graph and perform
$k+1$ link additions, one gets a $\cG_{N,0}$ graph, i.e. our original spin
glass problem with $N$ spins. Starting from the same $\cG_{N,2(k+1)}$ cavity
graph and performing $2$ site additions, one gets a $\cG_{N+2,0}$ graph, i.e.
our original spin glass problem with $N+2$ spins. Therefore the variation in
the GGS energy when going from $N$ to $N+2$ sites ($E_{N+2}-E_N$) is related
to the average energy shifts $\Delta E^{(1)}$ for a site addition, and $\Delta
E^{(2)}$ for a link addition, through:  
\be
E_{N+2}-E_N = 2 \Delta
E^{(1)}-{(k+1)}\Delta E^{(2)} \ .
\label{enetshift}
\ee
Using the fact that the total energy is asymptotically linear in $N$,
the energy density of the ground state 
is finally 
\be
U=\lim_{N \to \infty} {E_N}/N = {E_{N+2}-E_N \over 2}=\Delta E^{(1)}-{k+1 \over 2}\Delta E^{(2)} \ .
\label{enetot}
\ee

\begin{figure}[bt]
\centerline{    \epsfysize=5cm \epsffile{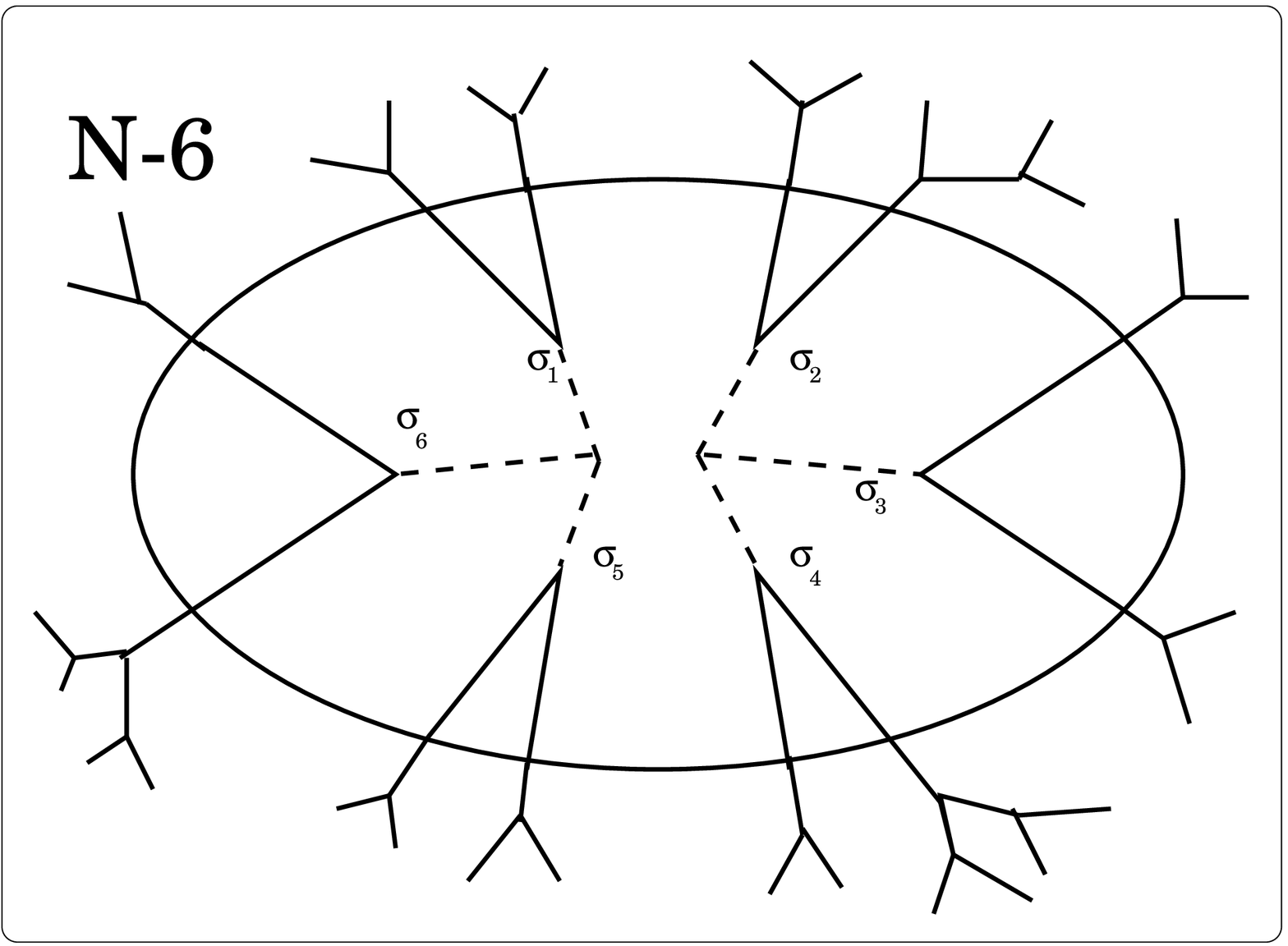} 
                \epsfysize=5cm \epsffile{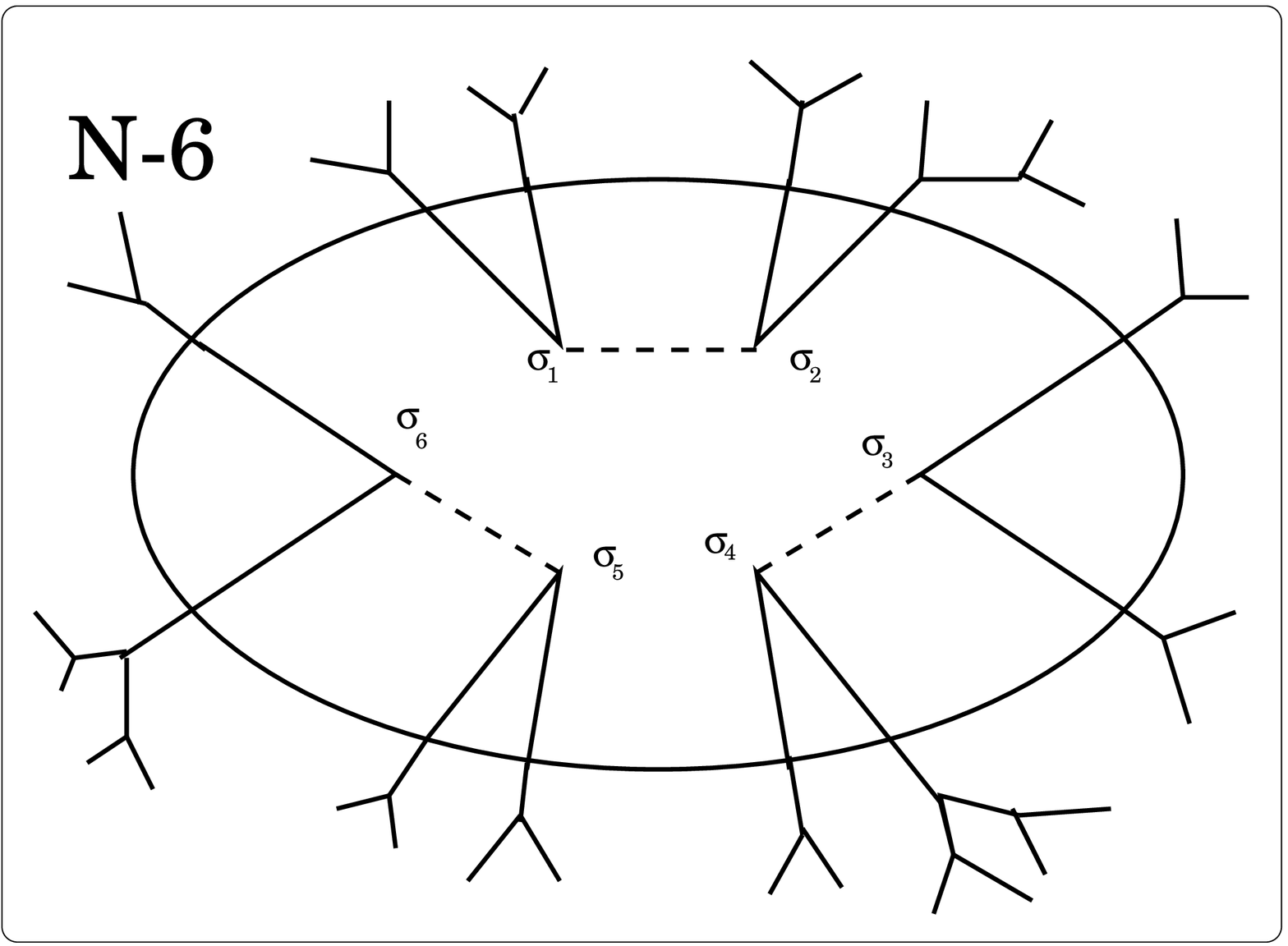}
}
\caption{Starting from the ${\cG}_{N,6}$ cavity graph,
one can either add two  sites (left figure) and create a 
${\cG}_{N+2,0}$ graph, or add three  links
(right figure) and create a 
${\cG}_{N,0}$ graph.}
\label{fignplusdeux}
\end{figure}

An intuitive interpretation of this result (for even $k+1$) is that in order
to go from $N$ to $N+1$ one should remove $(k+1)/2$ links (the energy for
removing a link is minus the energy for adding a link) and then add a site.
There are several alternative derivations and expressions of this energy
density \cite{fn1}, but the previous one is the simplest one for our purpose.

\section{``Replica symmetric'' solution}
\label{RSsect}
\subsection{General case}
When $q/N \ll 1$, generically, the distance on the lattice between two generic
cavity spins is large (it diverges logarithmically in the large $N$ limit). It
is thus reasonable to assume that the various cavity spins become
uncorrelated. This is the basic assumption of the RS solution,
although it is in general not explicitly formulated.
Several papers have worked out this RS solution in details at all temperatures
\cite{klein,katsura,nakanishi,bowman,thouless,chayes,mp86,kansom,carlsonetal}.
Here we are presenting for completeness 
the zero temperature version of this RS approach, the modifications induced 
in the 1RSB case will be explained below.
 This RS assumption amounts
to saying that the GGS energy of a $\cG_{N,q}$ spin glass can be written as an
additive function of the values of the $q$ cavity spins:  
\be E(\{\sigma
\})=E^0-\sum_{i=1}^q h_i \sigma_i \ . 
\ee 
The quantities $h_i$, that we call the
local cavity fields, depend on the sample. When considering the ensemble of
random cavity graphs, they are independent identically distributed (iid) random
variables, and their distribution is denoted $\cP(h)$. The computation of
$\cP(h)$ will be a crucial step in our approach. The reader should notice that
in general there is {\sl no} simple expression for the local  cavity fields
of the type: $ h_i\equiv\sum_{k}J_{ik}\sigma_{k} \ : $ These fields
 are related to the difference in energy of two GGS with
flipped cavity spins, and these two GGS may in principle differ in an
arbitrarily large number of spins. Indeed the quantity $E(\{\sigma \})$ is
computed by minimizing the energy as function of the other $N-q$ spins for
fixed values of the $q$ cavity spins.

\begin{figure}[bt]
\centerline{ \epsfysize=6cm \epsffile{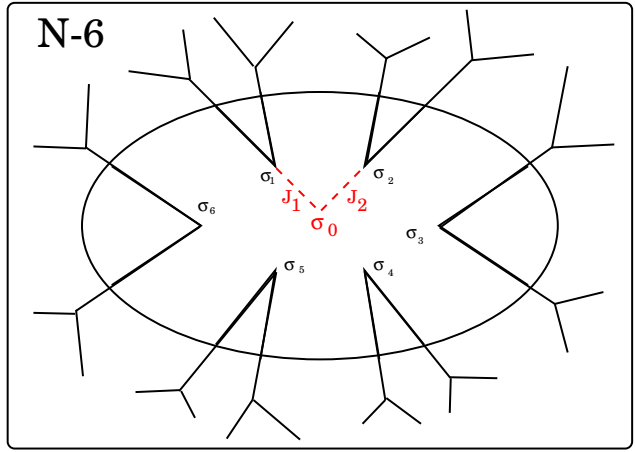}} 
\caption{In the iteration procedure, one adds 
a new spin into the cavity, and connects it to $k$ randomly chosen cavity spins, with some randomly 
chosen couplings.  Here an example with $k=2$.}
\label{iterate_fig}
\end{figure}

The key hypothesis of the RS treatment
is that the GGS of the spin glass before and after any of the previous graph
operations (e.g. iteration) are related. Equivalently one should assume that
the perturbation corresponding to the variation of one of the cavity spins
remains {\sl localized} and it does not propagate to the whole lattice. Under
this hypothesis (whose consistency can be checked and, as we shall see later,
is not always correct) a self consistent equation for this distribution is
easily found by considering the iteration procedure (see
fig.\ref{iterate_fig}). We add a new spin in the cavity, fix its value to
$\sigma_0$, and connect it to the spins $\s_1,...,\s_k$ through a new set of
coupling constants $J_1,...,J_k$. Each of the spins $\s_i$, $i \in \{ 1, ...,k
\}$, sees a local magnetic field $h_i+J_i \sigma_0$, and has to point into the
direction of this field in order to minimize its energy (if the local field is
zero this spin is free to point in any direction). The energy of the link
between $\s_0$ and $\s_i$ is thus: 
\be \eps_i=\min_{\sigma_{i}} (-h_i-J_i
\sigma_0) \sigma_{i}= - | h_i+J_i \sigma_0| \equiv -a(J_i,h_i)-\s_0 u(J_i,h_i)
\ , 
\ee 
where an elementary computation gives:
\be u(J_i,h_i)\equiv {1
\over 2} \(( \vert h_i+J_i\vert - \vert h_i-J_i\vert\)) \ \ , \ \
a(J_i,h_i)\equiv {1 \over 2} \(( \vert h_i+J_i\vert + \vert
h_i-J_i\vert\))=\vert h_i\vert +\vert J_i\vert- \vert u(J_i,h_i)\vert \ . 
\label{uadef}
\ee
This shows in particular that the new local field on site $0$, i.e. the
coefficient of $\s_0$ in the expression for the GGS energy, is given by
\cite{fn2}: 
\be h_0=\sum_{i=1}^k u(J_i,h_i) \ . 
\label{h0def}
\ee

This implies a   recursion relation for the distribution $\cP(h)$:
\be
\cP(h)= E_{J} \[[ \int \prod_{i=1}^k\[[ dh_i \cP(h_i)\]] 
\delta\((h-\sum_{i=1}^k u(J_i,h_i)\)) \]]
\label{Piter}
\ee 
(Throughout this paper, we use a generic notation $E_J[\cdot]$ to indicate
the average over the distribution of all coupling constants and over random
graphs).

Let us suppose that the previous equation is sufficient to 
determine the distribution
$\cP(h)$. If we know  this distribution we can compute  the energy.
The average energy shift due to a site addition is
\be
\Delta E^{(1)}=- E_{J}\[[ \int \prod_{i=1}^{k+1}\[[ dh_i \cP(h_i)\]]
\(( \sum_{j=1}^{k+1} a(J_j,h_j)+\vert \sum_{j=1}^{k+1} u(J_j,h_j)\vert\)) \]]\ ,
\label{e1rs}
\ee
and the average energy shift due to a link addition is
\be
\Delta E^{(2)}=- E_{J}\[[ \int \prod_{i=1}^{2}\[[ dh_i \cP(h_i)\]]
\max_{\s_1,\s_2} \((h_1 \s_1+h_2 \s_2 + J \s_1 \s_2\)) \]]\ .
\label{e2rs}
\ee
These expressions give the ground state energy 
density through (\ref{enetot}).

\subsection{Application to the $\pm$J model}

In order to obtain the RS solution, one must first solve the functional 
equation (\ref{Piter}) for $\cP(h)$.  
This is particularly easy in the case where the couplings are 
taken from the distribution $P(J)=(1/2)(\delta(J-1)+\delta(J+1))$. 
 The definition of the 
local fields implies that they are integers, and 
 the functions $u$ and $a$ take simpler forms:
\be
a(J,h)=\vert h \vert +\delta_{h,0} \ \ \ , \ \ \ u(J,h)=J S(h) \ ,
\ee
where $S(h)=0$ if $h=0$, $S(h)=\mbox{Sign}(h)$ otherwise. In other words the function $S(h)$
is $-1$ or $+1$ depending on the sign of $h$ and it is zero for $h=0$.

We can write the most general 
$\cP(h)$ in the form
\be
\cP(h)=\sum_{r=-k}^k p_r \delta(h-r) \ .
\label{simpleP}
\ee
Furthermore the symmetry of the $J$ distribution, together with (\ref{Piter}),
implies that $\cP(h)$ is symmetric ($p_r=p_{-r}$). 
The property of the function $S(h)$ implies that what matters is only the sign of $h$. The 
probabilities $p_{+}=p_{-}$ for having a positive  or negative $h$ are obviously given by 
$(1-p_{0})/2$, where in our notation  (\ref{simpleP}) $p_0$ is
probability of having a zero $h$. The self consistency
condition (\ref{Piter}) now becomes a closed 
equation for $p_0$,
\be
p_0=\sum_{q=0}^{[k/2]}
 C_k^{2q} 
p_0^{k-2q}\(({1-p_0\over 2}\))^{2q} 
 C_{2q}^q \ ,
\ee
(where $C_k^{q}={k! /(q! (q-k)! )   }$) 
and the other coefficients are given by
\be
p_r=p_{-r}=\sum_{q=0}^{[(k-r)/2]} 
 C_k^{2q+r} 
p_0^{k-2q-r}\(({1-p_0\over 2}\))^{2q+r} 
 C_{2q+r}^q \ .
\ee

If we specialize for simplicity to a Bethe lattice with $k=2$,
 the solution is $p_0=1/3,p_1=2/9,p_2=1/9$, and the ground state 
energy is $E=-23/18 \approx- 1.278$.
All these results were obtained in \cite{mp86}.

We would like to comment here on the fact that the fields are integers. It was
noticed already long ago that, even in the simple RS case, the iteration
equations (\ref{Piter}) admit some solutions which are not distributed only on
integers, but have a continuous part; in fact a lot of efforts have been
devoted to find 'the best' among these solutions. Moreover, if a computation
is done at finite temperature, where no ambiguity is present, and the results
are extrapolated at zero temperature, a continuous part is found in this RS
case. A detailed computation shows that if, working at zero temperature, one
starts from fields that are integer plus a small correction, this correction
is amplified under the iteration procedure until the final distribution is no
more concentrated near the integers. In our opinion it is quite likely that
this instability is a signal that the RS solution is incorrect. A similar
effect can be found on the 1RSB solution but it is weaker (see appendix A). We
believe that whenever one reaches the correct solution (which in the present
case should be full RSB), this artefact will disappear. Roughly speaking the
rational for this belief is the following: the fact that a small variation of
the field in one point propagates and leads to a large effect throughout the
whole lattice is precisely the so called ``replicon'' instability (that would
make the whole approach inconsistent). The RSB solution was invented just to
cure this problem and we know that it is very successful in doing this
\cite{MPV}.

\section{ Replica Symmetry breaking at the one step level} 
\label{1RSBsect}
\subsection{Basic hypotheses of the cavity method}
It is well known that the above result is wrong
\cite{motti,motti2,dewar,Lai,DDGold,MP_Bethe}, because the hypothesis of
continuity of the global ground state when we add new spins or links to the
graph is incorrect.

A mechanism  leading to the breaking of this continuity 
hypothesis, which is not taken 
into account in the RS approximation, is 
the existence of several local ground 
states (LGS, not to be confused with GGS).  By a LGS 
here we mean a state whose 
energy cannot be lowered by flipping a finite number of spins.  
Of course such a 
definition applies only to the $N=\infty$ limit.  In principle one 
should define the LGS 
for a finite system as being stable with respect to flipping a 
number of spins less than 
$f(N)$, where $f(N)$ is a well chosen increasing  function of $N$, 
diverging at large $N$. Determining the
 most appropriate 
form of $f$ is a difficult problem.  In its simplest version, 
the cavity method assumes the 
existence of a function $f(N)$ such that there are many ground states, 
but does not try to prove it.  Controlling this point would 
allow one to turn the cavity into a full 
mathematical proof, and has been done so far in only a few non trivial cases 
\cite{talag,bovier}).  Here we shall not try to provide this kind of 
rigorous treatment, we 
just want to show how the cavity method works, within its 
usual assumptions, and check it 
through the validity of its results. Some further comments on the
definition and counting of LGS are displayed in appendix C.
Although the precise definition of LGS is a very subtle issue,
we want to underline that there exist at least one model, rather closely
related to our Bethe lattice spin glass problem, where the present one step
RSB cavity method can be applied \cite{pspin}.
In this system the LGS can be characterized and enumerated fully 
 and the results can be checked versus
exact computations done using completely different methods
\cite{pspin,CDMM}.

In presence of several ground states, the assumption is that there is a one-to-one
 correspondence among the LGS before and after the addition of spins or
links (at least for the LGS with low energies). Equivalently we assume that
the perturbation due to the change of the value of a cavity spin propagates
(in the limit $N$ going to infinity) only to an infinitesimal fraction of the
lattice. Therefore it is possible to write an iteration procedure for the
whole population of LGS with given energy. However it may well be that the
order of the LGS energies change during the graph operations, and the GGS
after iteration is not the same LGS as the one before. The problem is to take
into account these 'level crossings', which is not done in the RS solution of
the previous section, and is done in the RSB solutions. Below we shall
describe one such solution, which is equivalent to what is called in replica
language the 1RSB solution. One is forced to follow a large population of the
LGS of lowest energy, large enough so that one can be sure to obtain the GGS
when iterating.

Let us state here the  postulates of the cavity method, at the level 
equivalent to the 1RSB solution in 
the replica language.
\begin{itemize}
\item The cavity spins are uncorrelated within one given LGS. Labelling by
$\alpha$ a LGS of a $\cG_{N,q}$ spin glass, its energy $E^\alpha$ is given by:
\be E(\{ \sigma \}) ^\alpha=E_0^\alpha-\sum_{i=1}^q h_i^\alpha \s_i \ . \ee
The previous equation would be wrong for the GGS because the GGS may
correspond to different LGS depending on the value of the cavity spins
$\sigma$. This extreme sensitivity of the GGS to the variation of a single
variable is typical of hard optimization problems.
\item
The energies $E_0^\alpha$ of the LGS of low energy (near to that of the GGS)
are assumed to be iid variables with a distribution given by a Poisson process
of density 
\be 
\rho(E_0)= \exp(\mu(E_0-E_{ref})) 
\ee 
where $E_{ref}$ is a
reference energy, which is near to the GGS energy, and $\mu$ is a parameter
whose physical interpretation  will be explained in sect. \ref{sect_sigma}.
 This
hypothesis is compatible with the idea that LGS are extremes of the energies,
and with the Gumbel universality class for extremes \cite{gumbel,jpbmm}.
\item
On a given site $i$, the local cavity fields in the various states,
$h_i^\alpha$, are iid variables taken from the same distribution $P_i(h)$.
However the distribution $P_i(h)$ fluctuates from site to site, so that 
the correct order parameter is a functional $\cQ[P(h)]$ giving the probability,
when one picks up a site at random to find on this site a cavity field distribution 
$P_i(h)=P(h)$ \cite{fn5}.
Moreover the cavity fields and the LGS energies are not correlated. 
\end{itemize}

These hypotheses are by no means evident and are likely to be wrong in many
models where subtle correlations exist among the different cavity field. This
more complex situation can appear, for instance, in cases where higher order
breaking of the replica symmetry is present and it will not be discussed here.
One should note that, while the 1RSB which we discuss here has a nice
interpretation in terms of LGS having independent energies (and thus a Gumbel
distribution), the higher order RSB solutions involve correlated variables and
thus describe new universality classes of extremes. In the spin glass case
which we study here, we expect that an infinite order of RSB will be needed to
solve the problem (as is well known for the large $k$ case \cite{parsol,MPV},
which is equivalent to the Sherrington Kirkpatrick (SK) model \cite{SK}). The
1RSB which we explore is thus an approximation to the problem, which usually
produces better results than the RS approach. In the limit where $k$ goes to
infinity it is know the the GGS energy per spin (divided by a normalization
factor $k^{-1/2}$) is -.798 in the RS case, -.766 in the 1RSB , to be compared
with the conjectured correct value -.763.
 
It is interesting to notice that the RS case of only one LGS can be obtained
in this framework by fixing the local distribution to be
$P_i(h)=\delta(h-a_i)$, so that all the LGS are automatically equal. On a
given site $i$, this distribution is fixed by the single number $a_i$. The
various $a_i$'s are iid, taken from a distribution $\cP(a)$ which satisfies
exactly the RS recursion relation (\ref{Piter}), and the result is $\mu$
independent, as it should. Alternatively in the limit where $\mu \to 0$, the
number of different LGS does no more increase with the energy, the gap between
the two lowest LGS diverges, and thus level crossings can be neglected. We
shall see that in this limit we again  recover the RS value of the average energy.

\subsection{Self consistency under iteration}
Let us study the effect of the iteration, starting from these hypotheses.
When iterating, the new local cavity field, in one given LGS  labeled by $\alpha$,
is given by the same expression as (\ref{h0def}):
\be
h_0^\alpha=\sum_{i=1}^k u(J_i,h_i^\alpha) 
\ee
and the energy of the LGS $\alpha$ is globally shifted from $E_0^\alpha$ to
$E_0^\alpha + \Delta E_0^\alpha$, where
\be
\Delta E_0^\alpha =- \sum_{i=1}^k a(J_i,h_i^\alpha) \ .
\ee 
The functions $u$ and $a$ are given in (\ref{uadef}).

In this case the  energy shifts and 
the local fields  on the new spin $\s_0$ are correlated. Given a
site (and thus fixing the couplings $J_i$), 
the pairs $(h_0^\alpha,\Delta E_0^\alpha)$ are iid, taken from a distribution
$P_{0}(h_0,\Delta E_0)$ which can be read-off from the iteration equations above:
\be
P_{0}(h_0,\Delta E_0)= \int \prod_{i=1}^k\[[ P_i(h_i) dh_i\]]
\delta\((h_0-\sum_{i=1}^k u(J_i,h_i)\))
\delta\((\Delta E_0^\alpha+\sum_{i=1}^k a(J_i,h_i)\))
\ee
Let us call ${E_0^\alpha}'= E_0^\alpha+\Delta E_0^\alpha$ the LGS energy after adding
the new spin $\s_0$. We first note that ${E_0^\alpha}'$ are iid, described again
by a Poisson process with an exponential density:
\be
\rho(E_0')= \exp(\mu (E_0-E_{ref}))\int dh_0 d (\Delta E_0)\;  P_{0}(h_0,\Delta E_0)
\exp(-\mu \Delta E_0) \ .
\label{eneshift_iter}
\ee
Therefore the exponential distribution is stable under iteration. 
The only effect of the iteration is a shift of the reference energy
from $E_{ref}$ to some new value $E_{ref}'$. 

As we are concerned with the LGS of lowest energies, we should study the 
states which have energy close to  $E_{ref}'$ (alternatively, if we
want to follow the population of the $M \gg1$ LGS with lowest energies, 
we need to keep only the
 states with the lowest {\it new} energy). 
 The joint distribution 
$R_{0}(h_0,E_0')$ of the local field and the new LGS  energy is given
by:
\begin{eqnarray} \nonumber
R_{0}(h_{0}, E_0')&\propto& \int dE_0 d (\Delta E_0) \exp(\mu (E_0-E_{ref}))
P_{0}(h_{0},\Delta E_0)\delta(E_0'-E_0-\Delta E_0)\\
&\propto&
\exp (\mu (E_0'-E_{ref}')) P_{0}'(h_{0})
\ ,
\end{eqnarray} 
where
\begin{eqnarray} \nonumber
 P_{0}'(h_{0}) &=&C \int d (\Delta E_0) \  P_{0}(h_{0},\Delta E_0)
\exp(-\mu \Delta E_0) \\ 
&=& C\int \prod_{i=1}^k P_i(h_i)
\delta\((h_0-\sum_{i=1}^k u(J_i,h_i)\))
\exp\((\mu\sum_{i=1}^k a(J_i,h_i)\))
\label{MAIN}
\end{eqnarray} 
the constant $C$ being fixed in such a way that $P_{0}'(h_{0})$ 
is a normalized probability 
distribution. 

The crucial point is the following: for each given LGS the recursion for the
probability distribution of the $h$'s is the naive one, i.e. the previous one
with $\mu=0$. However if we look at the recursion for the probability
distribution of the $h$'s, conditioned to a given value of the LGS energy we
find that it depends crucially on $\mu$. The probability distribution of the
cavity field in a generic LGS is different from the probability distribution of
the cavity field for a given value of the LGS energy. The two quantities may
be different because we are looking at the LGS of low energy which are for
large $N$ an infinitesimal fraction of the total number of local ground states
(indeed we shall see later that the total number of local ground states
increases exponentially with £$N$).

This distribution $P_{0}'$ of the   cavity field
on the new site depends on the added site through the choice of the
various $P_i$'s and of the coupling constants $J_i$. When one
averages over these choices, the distribution of $P_{0}'(h)$ must
be equal to the original $\cQ[P(h)]$: when  this self-consistency 
equation is satisfied, the hypotheses of the cavity method on the
$N$ spin system are found to be true again in the $N+1$ spin system 
obtained after iteration.

\subsection{Computation of the energy}
In order to compute the GGS energy, we must see the effect of 
adding a site or adding
a link within this 1RSB scenario.

{\it Site addition:} the energy shift during a site addition is computed
through the same method as (\ref{eneshift_iter}), 
the extra term coming from the fact that one optimizes the value of the added spin:
\be
\exp\((-\mu \Delta E^{(1)}\))= \int \prod_{i=1}^{k+1} \[[ dh_i P_i(h_i)\]]
\exp\((\mu  \sum_{i=1}^{k+1} a(J_i,h_i)
+\mu \vert \sum_{i=1}^{k+1} u(J_i,h_i)\vert \)) \ .
\label{e1_rsb}
\ee

{\it Link addition:}
\be
\exp\((-\mu \Delta E^{(2)}\))=  \int \prod_{i=1}^{2}\[[ dh_i P_i(h_i)\]]
\exp\(( \mu \  \max_{\s_1,\s_2} \((h_1 \s_1+h_2 \s_2 + J \s_1 \s_2\))\)) \ .
\label{e2_rsb}
\ee

Note that one gets back the RS expressions when  $\mu \to 0$.
Once the distribution $\cQ[P(h)]$ of the cavity field distributions is known,
one can compute the energy shifts $\Delta E_1, \Delta E_2$ which depend on $\mu$,
and deduce from them the energy function $\Phi(\mu)$ defined as in (\ref{enetot}):
\be
\Phi(\mu)= \Delta E^{(1)}-{k+1 \over 2}\Delta E^{(2)} \ .
\label{phidef}
\ee
 One finally finds:
\be
\Phi(\mu)= E_J\[[
\int \prod_{j=1}^{k+1}d\cQ[P_j]\; \Delta E^{(1)}[P_{1}\ldots P_{k+1}] \]]
-{k+1 \over 2}E_J\[[\int d\cQ[P_1]d\cQ[P_{2}]\Delta E^{(2)}[P_{1}\,P_{2}]
\]]
\label{Ufull0}
\ee
where $d\cQ[P]\equiv \cQ[P] dP$ denotes the probability distribution
of the probability $P$ (the integral is done over all the probability
distributions $P$ with weight $\cQ[P]$).
Explicitly, one has:
\begin{eqnarray} \nonumber
\Phi(\mu)&=& -{1 \over \mu} E_J\[[
\int\prod_{j=1}^{k+1}d\cQ[P_j]\;
\log\((
\int \prod_{i=1}^{k+1} \[[ dh_i P_i(h_i)\]]
\exp\((\mu  \sum_{i=1}^{k+1} a(J_i,h_i)
+\mu \vert \sum_{i=1}^{k+1} u(J_i,h_i)\vert \))
\))\]]
\\ \nonumber
&+&{k+1 \over 2 \mu}E_J\[[\int d\cQ[P_1]d\cQ[P_{2}] \right.\\
&&\left.\log\((
\int \prod_{i=1}^{2}\[[ dh_i P_i(h_i)\]]
\exp\(( \mu \  \max_{\s_1,\s_2} \((h_1 \s_1+h_2 \s_2 + J \s_1 \s_2\))\))
\))\]]
\label{Ufull}
\end{eqnarray} 
Notice that all this program can be carried out for any given $\mu$. In order
to fix the parameter $\mu$ which gives the increase rate of the number of LGS,
one must compute $\Phi$ as a function of $\mu$ and maximize it. One
justification for that procedure can be to go back to the replica formulation:
$\mu$ turns out to be the zero temperature limit of $m/T$, where $m$ is the
breakpoint in Parisi's order parameter function at the 1RSB level \cite{MPV}.
The necessity of maximising over $m$ is a well known feature of the replica
method (in the limiting case $k \to \infty$ it can be rigorously proven
\cite{guerra} that the value of $\Phi(\mu)$ is a {\sl lower} bound to the
correct result so that it is natural that the preferred value of $\mu$ is
obtained by maximizing $\Phi(\mu)$), but this does not provide a clear
physical reason.

The next section will explain the physical meaning of $\mu$, and explain why
$\mu$ must be equal to $\mu^*$, chosen such that $\Phi(\mu)$ is maximum, in
order to compute the GGS energy.

It is also interesting to notice that the whole self-consistency procedure
obtained by the iteration equation (\ref{MAIN}) is a variational procedure
which can be deduced from the stationarity condition of the functional
$\Phi(\mu)$ given in (\ref{Ufull}) with respect to changes of ${\cal Q} [P]$.
This is shown in appendix B. The existence of this variational formulation is
crucial in simplifying the computation of all the derivatives with respect to
the various parameters, because only explicit derivatives must be considered.

\subsection{Computing the complexity}
\label{sect_sigma}
In presence of many LGS,  one can be interested in knowing their
number.
We shall make the basic assumption that for large $N$, the 
typical
number $\cN_N(E)$
of LGS with a given energy $E$ behaves as
\be
\log(\cN_N(E)) 
\simeq N \Sigma\((\frac{E}{N}\)) \ .
\label{scaling}
\ee 
The function $\Sigma(\eps)$ is a positive function
called the complexity of the problem. 
In the range of energy densities $\eps$ where $\Sigma(\eps)>0$, the $\log$ of
the number of LGS is supposed to be a selfaveraging function, so that for
almost all sample (in the large $N$ limit), $(1/N) \log(\cN_N(E))$ is given by
$\Sigma$. Technically, given a definition of the LGS (see the discussion in
app. C), one can thus define the complexity as $\Sigma(\eps)= \lim_{N \to
\infty} (1/N) E_J \log(1+\cN_N(N\eps))$.
In general, for ground states $\alpha$ with a fixed energy density
$\eps=E^\alpha/N$, the local cavity fields $h_i^\alpha$ are iid variables
taken from the distribution $P^{(\eps)}_i(h)$, which now depends on $\eps$,
and fluctuates from site to site. The corresponding order parameter is a
functional $\cQ^{(\eps)}[P]$ giving the probability, when one picks up a site
at random to find on this site a cavity field distribution (for states with
energy density $\eps$): $P^{(\eps)}_i(h)=P(h)$.

Our goal here is to check the self consistency of these hypotheses under
iteration, and to determine the 'complexity' function $\Sigma$. One expects
that $\Sigma(\eps)$ will vanish at some value $\eps_0$, which gives the GGS
energy density, and the analysis of the previous section applies for LGS
having energy close to $N \eps_0$ (plus terms of order one). In this context
the value of $U$ can be found by looking for the solution of the equation \be
\Sigma(U)=0 \ . \ee

Let us first study the effect of the iteration.
The number of LGS at energy $E$ after iteration is given by
\be
\cN_{N+1}(E)=\int P_{0}(h_0,\Delta E) dh_0 d \Delta E
\exp\((N \Sigma\(({E-\Delta E \over N}\))\)) \ .
\label{nnp1}
\ee
We need to compute the distribution of local cavity fields
$P_{0}^{(\eps)}(h)$ at a fixed GS energy density $E/N=\eps$.
Expanding to first order in the small shift $\Delta E/N$ in
(\ref{nnp1}), we obtain:
\bea
P_{0}^{(\eps)}(h)&=&C \int P_{0}(h_0,\Delta E)d \Delta E
\exp\(( -\mu \Delta E\))\\
&=& C \int \prod_{i=1}^k \[[ P_i^{(\eps)}(h_i) dh_i\]]
\delta\((h_0-\sum_{i=1}^k u(J_i,h_i)\))
\exp\((\mu\sum_{i=1}^k a(J_i,h_i)\))
 \ ,
\label{reweighting_eps}
\eea
where
\be
\mu={d \Sigma (\eps) \over d \eps}\ .
\label{mudef}
\ee
 Notice that we get back exactly the same expression as in the
previous section (see(\ref{MAIN})), but now $\mu$ has 
a well defined meaning: it is the derivative of
the complexity with respect to the energy density $\eps$. When one varies
$\eps$, the value of $\mu$ changes. The whole formalism of the previous section
can be used for any $\mu$. Its result will give the distribution of
cavity fields for the states $\alpha$ which have  energy density $\eps$,
which is related to $\mu$ through (\ref{mudef}). The
functional probability distribution found in this way is  $\cQ[P^{(\eps)}(h)]$.

In order to compute the complexity function $\Sigma(\eps)$, 
we must see the effect of adding $2$ sites or adding
$k+1$ links to a $\cG_{N,2k+2}$ graph within this 1RSB scenario.
Let us call $\Sigma_{2k+2}(\eps)$ the complexity function of the
$\cG_{N,2k+2}$ graph from which we start. 

{\it Site addition:}
 By adding one site,
we go to a $\cG_{N+1,k+1}$ graph with complexity $\Sigma_{k+1}(\eps)$.
To compute this, we notice that 
when   one site is added and the corresponding spin is optimised,
 the energy of the LGS is shifted by a value  $\Delta E$, and the
corresponding probability distribution $P_{site}(\Delta E)$ is:
\be
P_{site}(\Delta E)=\int \prod_{i=1}^{k+1} \[[ dh_i P_i(h_i)\]]
\delta\((\Delta E+  \sum_{i=1}^{k+1} a(J_i,h_i)
+ \vert \sum_{i=1}^{k+1} u(J_i,h_i)\vert \)) \ .
\ee
After the site addition, the  new 
complexity $\Sigma_{k+1} (\eps)$ is:
\begin{eqnarray} \nonumber
\exp\[[(N+1) \Sigma_{k+1}\(( {E \over N+1}\))\]]
&=&
\int P_{site}(\Delta E) d \Delta E \ \exp\[[N \Sigma_{2k+2}\(( {E-\Delta E \over N}\))\]]
\\
&=&
\exp\[[N \Sigma_{2k+2}\(( {E}\))\]]
\int P_{site}(\Delta E) d \Delta E \exp\[[-\mu \Delta E\]]
\end{eqnarray} 

{\it Link addition:}
By adding one link,
we go to a $\cG_{N,2k}$ graph with complexity $\Sigma_{2k}(\eps)$.
The probability distribution for the corresponding 
energy shift is:
\be
P_{link}(\Delta E)= \int \prod_{i=1}^{2}\[[ dh_i P_i(h_i)\]]
\delta\((\Delta E+ \  \max_{\s_1,\s_2} \((h_1 \s_1+h_2 \s_2 + J \s_1 \s_2\))\)) \ .
\label{e2_link}
\ee
After the link addition, the  new 
complexity $\Sigma_{2k} (\eps)$ is:
\begin{eqnarray} \nonumber
\exp\[[N \Sigma_{2k}\(( {E \over N}\))\]]
&=&
\int P_{link}(\Delta E) d \Delta E \ \exp\[[N \Sigma_{2k+2}\(( {E-\Delta E \over N}\))\]]
\\
&=&
\exp\[[N \Sigma_{2k+2}\(( {E}\))\]]
\int P_{link}(\Delta E) d \Delta E \exp\[[-\mu \Delta E\]] \ .
\end{eqnarray} 

{\it Going from a $N$ site graph to a $N+2$ site graph:}
As before, we need to add two sites and take away $k+1$ links.
When performing two times the site addition, we go to a graph 
$\cG_{N+2,0}$ where the number of GS is  
\be
\exp\[[ (N+2)\Sigma\(({E\over N+2}\))\]]\simeq\exp\[[ N\Sigma\(({E\over N}\))\]]
 \exp\[[ 2\Sigma\(({E\over N}\))
- 2 \(({E\over N}\))\Sigma'\(({E\over N}\))\]]
\ ,
\ee
while  performing
$k+1$ times the link addition leads to  a graph 
$\cG_{N,0}$ where the number of LGS is  $\exp\(( N\Sigma(E/N)\))$.
Adding up the effect of the link additions and the site additions,
one obtains:
\bea
\Sigma(\eps)-\eps\Sigma'(\eps)&=& 
E_J
\log\((\int P_{site}(\Delta E) d \Delta E \exp\((-\mu \Delta E\))\))\\
&&
-{k+1 \over 2}
E_J
\log\((\int P_{link}(\Delta E) d \Delta E \exp\((-\mu \Delta E\))\))
\label{final}
\eea
where the average $E_J$ also implies an averaging over 
the choices of  $P_i(h_i)$ appearing in $P_{site}$
and $P_{link}$ (the functions $P_i$ have to be taken from the
functional distribution $\cQ[P(h)] $).

Comparing to the expression for the energy density derived previously (\ref{Ufull}),
one gets simply:
\be
\Sigma(\eps)-\eps\mu=-\mu \Phi(\mu) \  , \ \ \mbox{where} \ \ \mu=\Sigma'(\eps) \ .
\label{legendre}
\ee
 So the knowledge of the function
$\Phi(\mu)$ allows to reconstruct the complexity function $\Sigma(\eps)$ through a
Legendre transform \cite{monasson}.
The usual relations for such a transform,
\be
{d (\mu \Phi)  \over d \mu}=\eps \ \ \ ,
\ \ \ {d  \Phi  \over d \mu}={\Sigma(\eps) \over \mu^2} = {\eps- \Phi(\mu) \over \mu},
\ee
show that the condition of maximisation of $\Phi(\mu)$
is equivalent to having $\Sigma=0$. This proves that the GGS energy density 
is found by maximising $\Phi(\mu)$ with respect to $\mu$. It also gives the practical
way of deducing the complexity from the knowledge of $\Phi(\mu)$. 
Because of the structure of (\ref{legendre}), we call $\Phi(\mu)$ the
zero temperature free energy of the problem. In fact
 the previous equations could also be written as
\be
\exp (-N \mu \Phi(\mu))\sim \int d\eps \exp\(( -N \eps \mu +N\Sigma(\eps)\))
\sim
\sum_{\alpha} \exp\((-\mu E_{\alpha}\)) \ ,
\ee
where $\alpha$ labels the LGS and $E_{\alpha}$ is the total energy of the LGS.
In other words $\Phi(\mu)$ is the free energy (at a temperature $1/\mu$) of a
system where the sum is done only on the LGS and not over all the spin
configuration (as in the case of the usual free energy). One could also apply
the same techniques used at finite temperature \cite{MP_Bethe} to rederive the
previous expression for $\Phi(\mu)$ and $\Sigma(\mu)$ in a slightly different way.

Let us give here for completeness the full expressions used in
the computation of the complexity. Once one has solved the iteration equations 
at a given $\mu$, and obtained the functional probability distribution $\cQ[P]$,
one can deduce $\Phi(\mu)$ from (\ref{Ufull}).
The energy density $\eps$ can then be obtained as:
\begin{eqnarray} \nonumber
\eps&=& - E_J \[[\int \prod_{j=1}^{k+1}d \cQ[P_j] 
{
\int \prod_{i=1}^{k+1} \[[ dh_i P_i(h_i)\]]
D_\mu\(( \sum_{i=1}^{k+1} a(J_i,h_i)+
\vert \sum_{i=1}^{k+1} u(J_i,h_i)\vert \))
\over
\int \prod_{i=1}^{k+1} \[[ dh_i P_i(h_i)\]]
\exp\((\mu  \sum_{i=1}^{k+1} a(J_i,h_i)+
\vert \sum_{i=1}^{k+1} u(J_i,h_i)\vert \))
} \]]
\\
&+&{k+1 \over 2}E_J\[[\int d\cQ[P_1] d\cQ[P_{2}]
{
\int \prod_{i=1}^{2}\[[ dh_i P_i(h_i)\]]
D_\mu\(( \max_{\s_1,\s_2} \((h_1 \s_1+h_2 \s_2 + J \s_1 \s_2\))\))
\over
\int \prod_{i=1}^{2}\[[ dh_i P_i(h_i)\]]
\exp\(( \mu \max_{\s_1,\s_2} \((h_1 \s_1+h_2 \s_2 + J \s_1 \s_2\))\))
} \]]
\label{epsfull}
\end{eqnarray} 
 where  we have defined the function
$D_\mu(x)=x e^{\mu x}$. The previous equations can be rewritten with an
evident meaning of the symbols as:

\begin{eqnarray} \nonumber
\eps&=& - E_J \[[\int \prod_{j=1}^{k+1}d \cQ[P_j]
{
\int \prod_{i=1}^{k+1} \[[ dh_i P_i(h_i)\]]
\Delta E^{1} \exp(- \mu \Delta E^{1} )
\over
\int \prod_{i=1}^{k+1} \[[ dh_i P_i(h_i)\]]
\exp\((-\mu   \Delta E^{1} \))
} \]]
\\
&+&{k+1 \over 2}E_J\[[\int d \cQ[P_1]d \cQ[P_{2}]
{
\int \prod_{i=1}^{2}\[[ dh_i P_i(h_i)\]]
\Delta E^{2} \exp(- \mu \Delta E^{2} )
\over
\int \prod_{i=1}^{2}\[[ dh_i P_i(h_i)\]]
\exp(- \mu \Delta E^{2} )
} \]]
\label{epsfullsimple}
\end{eqnarray} 
Finally the complexity is
\be
\Sigma(\eps)=\mu(\eps-\Phi) \ ,
\ee
where $\Phi$ is given in (\ref{Ufull}) and $\eps$ is given in (\ref{epsfull}).

The smart reader may cast some doubts on the consistency of the whole
approach: our first hypothesis was the existence of a one-to-one
correspondence between the ground states when increasing $N$ by two units (and
therefore the number of ground states should no depend on $N$) and from this
hypothesis we arrive to the conclusion that the number of ground states
increase with $N$ as $\exp (N\Sigma(\eps)) $. We believe that the hypothesis
of one-to-one correspondence of the LGS is valid only in a low energy region,
where the computation that we have presented is correct, and it should be
modified in the region where $\Sigma(\eps)$ is near to its maximum.

\subsection{Factorized case}
A particularly simple case is when we assume that the distribution $P_i(h)$
are $i$ independent (i.e. the $\cQ[P(h)]$ is a functional $\delta$ function).
This case, which was first studied in \cite{WonShe}, and developed for the
Bethe lattice spin glass in \cite{gollai}, is named the 'factorized case'
because of the special pattern of RSB to which it leads. It is simple because
the order parameter is a single function $P(h)$ (fixed from the
self-consistency relation (\ref{MAIN}) by imposing
$P_0'(h)=P_1(h)=P_2(h)=...=P_k(h)$), and the RS equation are only slightly
modified. However one should note that, in general, one may expect a $P_i(h)$
which fluctuates: this is obviously the case whenever the connectivity
fluctuates,
 but this may also happen
in the case of the fixed connectivity random graphs which we study
here. Some special models where the
factorized Ansatz gives an exact solution have been studied recently
\cite{montanari_codes,FLRZ}.

\subsection{Application to the $\pm$J model}
\subsubsection{Factorized case}
Our first task is to compute the  distribution $\cQ[P(h)]$.  
We will do it here first  in the simple factorized case where $P_i(h)$
does not depend on the site (so that $\cQ[P(h)]$
is a functional $\delta$).  The distribution $P_i(h)\equiv P(h)$
still has the form of a sum of delta peaks on the 
integers $P(h)=\sum_{r} p_r \delta(h-r)$. This factorized solution can be stable 
only if $P(h)$ is symmetric ($p_r=p_{-r}$).  
 The equations for the weights of the $\delta$ peaks 
$p_0,p_1,...,p_k$, taking into account the free energy shift, are:
\be
p_r=A\sum_{h_1=-k}^k...\sum_{h_k=-k}^k \delta\((r-\sum_{i=1}^k S(h_i)\)) p_{h_1}
...p_{h_k} \exp\((\mu\[[\sum_{i=1}^k (\vert h_i\vert +\delta_{h_i,0})\]]\))
\label{iterfacto}
\ee 
(the function $S$ has already been defined as $S(h)=0$ if $h=0$,
$S(h)=Sign(h)$ otherwise). The quantity $A$ is a normalisation constant fixed
by the condition 
\be 
1=p_0+2\sum_{r=1}^k p_r 
\ee 
One can notice the similarity
of (\ref{iterfacto}) with the RS solution (\ref{simpleP}). With respect to
this RS solution, the self consistency equations are only modified by the
reweighting term in $\exp(\mu...)$.
 However the physical interpretation of $P(h)$ studied
here and the $\cP(h)$ of the RS solution are totally different.
The present  $P(h)$  gives, for one site $i$, the probability that 
in a LGS $\alpha$
chosen at random, with energy density $\eps$ fixed  by the value of $\mu$,
the field $h_i^\alpha$ is equal to $h$. In the RS case, there is only 
one LGS, and $\cP(h)$ governs the fluctuation of $h_i$ from site to site.

In the case $k=2$, one  obtains, for any $\mu$, an algebraic solution:
\be
p_0=(24+3 e^{2 \mu} -(8+e^{\mu}) \sqrt{8+e^{2 \mu}})/(4 (e^{\mu}-1)^2) \ \ , \ \ 
p_1={2 p_0 q \over (p_0+2q)^2} \ \ , \ \ 
p_2={q^2 \over (p_0+2q)^2}
\label{solfact}
\ee
where 
\be
q={p_0 (3 p_0-2) \over e^{\mu} -4 p_0 -2 p_0 e^{\mu}}
\ee

\begin{figure}[t]
\begin{center}
\epsfig{figure=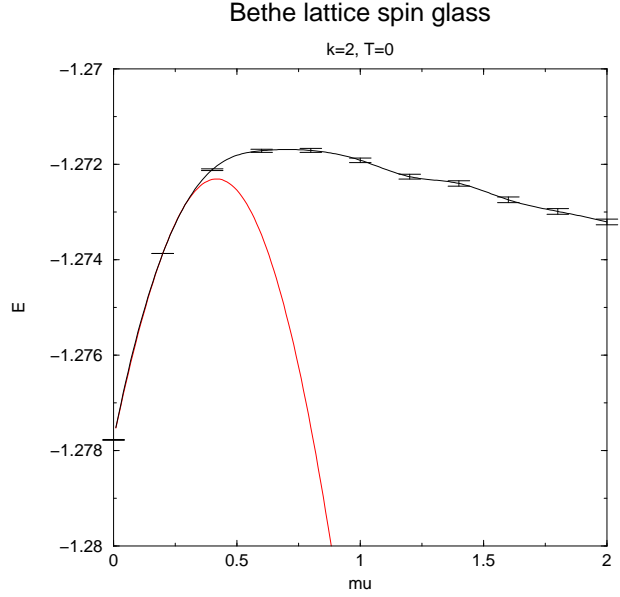,width=8cm,angle=-90}
\caption{The  {\it zero $T$  free energy } per site $\Phi(\mu)$ as a function of $\mu=\beta m$ for $k=2$,
$T=0$.
Shown are the  one step solution 
in the factorized case 
(dotted) and the correct 1RSB solution 
(full) 
computed by $10^4$ iterations of
a population of $10^3$(1RSB) or $10^4$(factorized) local probability distributions.
In the 1RSB solution, the full line is a spline interpolation 
through the data-points.
There is a clear evidence for a non-factorized solution, 
although the effect on the energy is small.
\label{fig_free}}
\end{center}
\end{figure}

After some algebra one finds that the energy shifts are given by:
\begin{eqnarray} \nonumber
\exp\((-\mu \Delta E^{(1)}\))&=& e^{3\mu}\[[p_0^3 +6(p_1+p_2 e^{\mu})\]]
+e^{4\mu}\[[3 p_0^2 (p_1+p_2 e^{\mu}) +3(p_1+p_2)^3\]]\\
&+&e^{5\mu}\[[3 p_0 (p_1+p_2 e^{\mu})^2 \]]
+e^{6\mu}\[[(p_1+p_2 e^{\mu})^3 \]]\\
\exp\((-\mu \Delta E^{(2)}\))&=&e^{\mu } \(( \[[2 p_1 e^{\mu }+2 p_2 e^{2 \mu }\]]
\[[ p_2 e^{2 \mu }+  p_1 e^{\mu } +p_0+ p_1 e^{-\mu }+p_2\]]+
p_0\[[2 p_2 e^{2 \mu }+ 2 p_1 e^{\mu } +p_0\]]\))
\end{eqnarray} 
Using the values of $p_0,p_1,p_2$ in (\ref{solfact}), we  obtain
from(\ref{Ufull})  the zero $T$ free  energy $\Phi(\mu)$
which is plotted in fig. \ref{fig_free}.
 This quantity  has a maximum at 
$\mu \simeq .4174$, where the GGS energy is equal to $E=-1.27231$.
All these results using the factorized Ansatz were derived 
previously with the replica approach in \cite{gollai}.

\subsubsection{The full 1RSB solution}
Let us now go beyond the factorized approximation. On a given site $j$ 
the probability distribution  of the cavity field is 
\be
P(h| \vp^j)=\sum_{r=-k}^k p_r^j \delta(h-r) \ .
\label{pform}
\ee
It is parametrized by a  vector of weights,  $\vp^j=(p^j_{-k},...,p^j_{k})$ 
which can fluctuate from one site $j$  to the next.
Because we work at zero temperature and the fields take  integer values, 
the probability
depends on a finite number $(2k)$ of parameters, and the
full order parameter is not a functional, but a function $R(\vp)$ of the vector of weights $\vp=(p_{-k},...,p_k)$, which is given 
in the limit of large $N$ by: 
\be
R(\vp)=\frac{1}{N} \sum_j \[[ \prod_{r=-k}^k \delta \((p^j_r-p_r\))\]] \ .
\ee

Let us perform one iteration, by adding a spin
$\s_0$ in the cavity,  connecting it to
 the k cavity sites $j_1,...,j_k$, and optimizing the
values of $\s_{j_1},...,\s_{j_k}$. The distribution of the cavity field on site 
$0$ again takes the form (\ref{pform}), with:
\be
p_r^0=A^0\sum_{h_{1}=-k}^k...\sum_{h_{k}=-k}^k 
\delta\((r-\sum_{i=1}^k S(h_i)\)) p_{h_1}^{j_1}
...p_{h_k}^{j_k} \exp\((\mu\[[\sum_{i=1}^k (\vert h_i\vert +\delta_{h_i,0})\]]\))
\ ,
\label{iterfull}
\ee
where $A^0$ is a site-dependent normalisation constant, fixed by $\sum_r p_r^0=1$.

The energy shifts are given by (\ref{e1_rsb}) and (\ref{e2_rsb}),
and are also site dependent.

The iteration (\ref{iterfull}) offers a way to obtain the order parameter
$R(\vp)$ by a  method of population dynamics \cite{MP_Bethe}. The method amounts
to following a population of $\cN$ vectors $\vp^1,...,\vp^{\cN}$. 
Each iteration of the dynamics corresponds precisely to the iteration of the cavity
method: one picks up $k$ vectors $\vp^{j_1},..., \vp^{j_k}$ at random
in the population, one computes the new vector $\vp^0$ according to 
(\ref{iterfull}), and substitutes the new factor $\vp^0$ in the population, 
in the place of a randomly chosen vector $\vp^i$. This stochastic process
yields, after some transient regime, a population whose distribution is stationary, 
and the probability distribution of $\vp$ inside this population is
nothing but the $R(\vp)$ order parameter.
The  energy shifts $\Delta E^{(1)}$ and $\Delta E^{(2)}$ obtained 
when adding one site or one link are computed as before, 
and they are then averaged over many iterations. This allows to compute,
for any $\mu$,
the zero $T$ free energy $\Phi(\mu)$ using (\ref{phidef}), which is 
 shown in fig. \ref{fig_free}. It has a maximum
at a value $\mu\sim.7$, and predicts a GGS energy density $U \simeq -1.2717$.

One should notice that in this problem the quantitative effect of RSB on the 
GGS energy density are rather small. The full 1RSB result  $U \simeq -1.2717$
differs from the RS result $U \simeq -1.2777$ by  $5. 10^{-3}$, and the
factorized solution gives a very good approximation $U\simeq-1.2723$,
precise at $5. 10^{-4}$. Therefore one can expect that the quantitative effects of
higher order RSB on the GGS energy will be tiny.

\begin{figure}[bt]
\begin{center}
\epsfig{figure=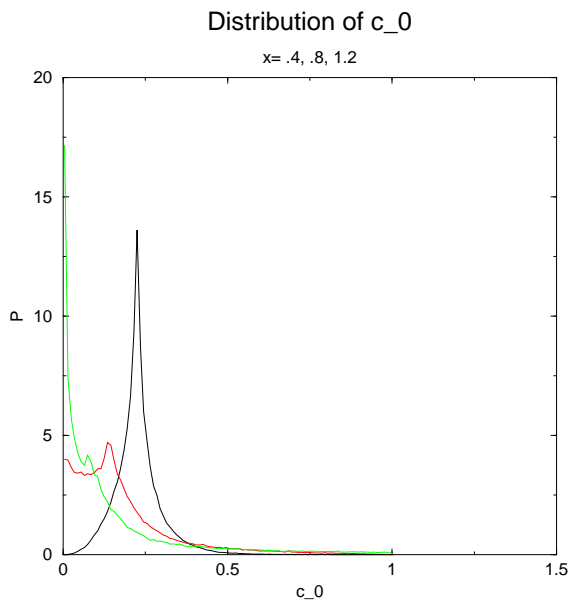,width=8cm,angle=-90}
\caption{Probability distribution of $p_0$, obtained after evolving
a population of $\cN=10^5$ sites. Plotted are the cases $\mu=.4$ (black),
$\mu=.8$ 
(grey) and $\mu=1.2$ 
(light grey). Notice the big effect of non-factorization.
The best factorized solution, with $\mu=.4174$,
 would give a $\delta$ peak at $p_0=.3353$. The RS solution would give a
$\delta$ peak at $p_0=1/3$.
\label{fig_c0}}
\end{center}
\end{figure}

However, the qualitative consequences of RSB  are clearly visible on the
fact that the probability distribution of cavity fields is site dependent.
This is exemplified by fig. \ref{fig_c0}, which gives the site to
site fluctuations
$P(p_0)=(1/N)\sum_i\delta \(( p_0^i-p_0\))$ of the probability of 
having a zero cavity field. We have also checked that, while the
individual cavity field distributions $P_j(\vp)$ are not symmetric 
under field reversal
(i.e. $P_j(p_r)\ne P_j(p_{-r})$), the full order parameter is
statistically symmetric (i.e. the site to
site fluctuations of $p_r$ are identical to those of $p_{-r}$).

\begin{figure}[bt]
\begin{center}
\epsfig{figure=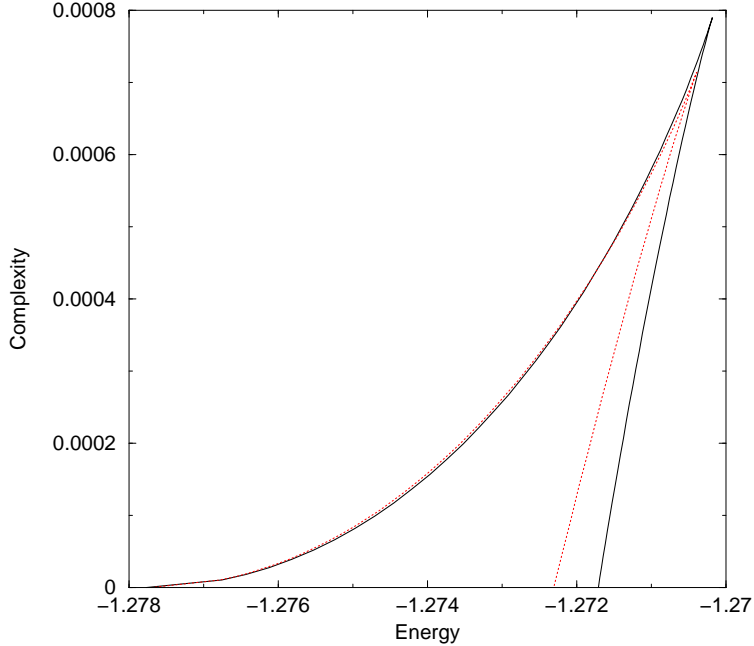,angle=-90,width=10cm}
\caption{The complexity $\Sigma$ as function of $\eps$, for the $k=2$ spin
glass on the Bethe lattice.
The full 
curve is the result of the full 1RSB solution,
the dotted 
curve is the result of the
 factorized approximation.
\label{comp_fig}}
\end{center}
\end{figure}

Using the previous results one can compute the complexity function $\Sigma(\eps)$.
In fig.\ref{comp_fig} we show the results in the region 
where the complexity is positive. 
In order to obtain these results we only need 
 the function $\Phi(\mu)$  in 
the region to the left of its maximum, where $d\Phi/d\mu \ge 0$. 

The complexity is rather similar in the factorized method and in the true solution, apart from some 
small variation of the ground state energy.  However it shows a somewhat unusual form which we wish 
to comment.  One sees two branches.  The right branch is concave and goes from $\eps=U$, the ground 
state energy where the complexity vanishes, to $\eps=\eps_M$, the maximal energy beyond which one 
does not find any local ground state, which corresponds to a value $\mu=\mu_M$.  It is obtained for 
$\mu \in [\mu_M,\mu^*]$, where $\mu^*$ is the point where $\Phi(\mu)$ reaches its maximum, 
corresponding to the GGS energy $U$.  The second branch is convex, and interpolates between the RS 
solution (obtained at $\mu=0$) and the maximal complexity point (obtained at $\mu=\mu_M$).  This 
second branch does not seem to have a direct physical interpretation and in this context can be 
simply ignored.
On the other hand it must be present insofar as the $\mu\to 0$ limit
of our RSB solution gives back the RS solution. 
Clearly a better understanding of this second branch would be welcome.

\begin{figure}[bt]
\begin{center}
\epsfig{figure=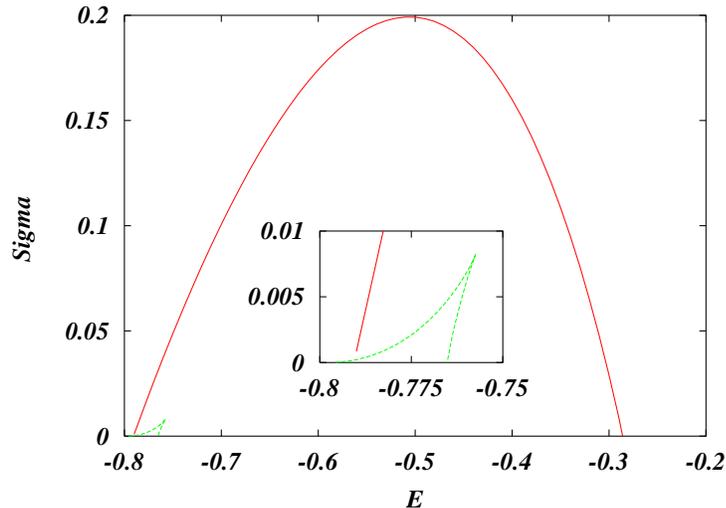,width=10cm}
\caption{The complexity $\Sigma$ as function of $\eps$, for the
Sherrington Kirkpatrick model
The top
curve is the complexity of configurations stable
to one spin flip, the 
bottom curve is the result of the 1RSB solution.
The inset shows the details of the curves near to the ground state energy.
\label{SK_fig}}
\end{center}
\end{figure}

It is curious that this type of complexity curve has not been discussed in the
literature so far, to our knowledge. In the rest of this section we want to
present a short comment on some apparent discrepancy between our results and
some widely used results on the SK model \cite{SK}. In the SK model which
corresponds to the $k \to \infty$ limit of our problem, the states can be
defined at any temperature as solutions of TAP equations \cite{TAP,MPV}. Bray
and Moore \cite{BM} have computed the number of such solutions, and they have
deduced that in the zero temperature limit, the corresponding complexity
becomes equal to the complexity of one spin flip stable (1SF) states, which
has been computed in \cite{BM,DDGGO}. In fig.\ref{SK_fig} we show the
complexity of 1SF states, and also the complexity at the 1RSB level computed
through the use of the Legendre transform of the free energy as in
(\ref{legendre}). This 1RSB result is indeed very different, qualitatively and
quantitatively, from the complexity of 1SF states, and also different from the
general shape found by Bray and Moore at finite temperatures.

It seems to us  that this difference can be explained by two
arguments: - The 1SF configurations are generically not stable to several spin
flips (in contradiction to what is claimed in \cite{BirMon}) and in particular
a change on the cavity fields may propagate through the whole system - The free
energy functions of \cite{BM} admit also other new saddle points which have
not yet been considered. Clearly a full discussion of these points goes much
beyond the scope of the present paper; much more work is needed in order to
understand the relationship between TAP states, the $q$-spin flip stable
states, and the Legendre transform construction in spin glasses\cite{MP_prep}.

\section{Conclusion} 
\label{conclsect}
With respect to our previous work \cite{MP_Bethe} on the cavity method for
the Bethe lattice spin glass, the fact of working directly at zero temperature
allows to simplify the discussion of the cavity method in several aspects
\footnote{We warn the maybe puzzled reader that the form of the free energy functional we use is 
different from that of our previous work \cite{MP_Bethe}. There are many equivalent 
free energy functionals that one can use that coincide when the recursion equations are 
satisfied and are variational. The one used in this paper has been selected because of the 
simplicity of the derivation (incidentally it leads to smaller statistical errors when 
doing numerical studies).}.
On
the one hand, one can discuss the physics directly in terms of ground states.
The physics of RSB comes into play when there exist many LGS and their
energies may cross when one adds a new spin. At a more technical level, the
cavity fields are integers, and it is thus much simpler to parametrize their
probability distributions. This allows in some cases to get analytic
solutions, but also when one has to resort to the numerical solution of the
cavity equations using the population dynamics, this aspects simplifies a lot
the procedure: one can follow a population of the local distributions
$P_i(h)$, because each of them is easily parametrized, while in the finite $T$
approach one must represent each $P_i(h)$ by a population of $\cal M$ fields.
Computationally, this zero $T$ approach is thus much more efficient. This is
very interesting for discussing optimization problems, and in fact it has
already allowed for a full solution of the Ksat problem \cite{Ksat}. We expect
that it can have many other applications in this context.

\section*{Appendix A: The $T\to 0$ limit and the RSB level}
If we study the thermodynamics of the model at finite temperature, we can use a very 
similar approach where the free energy of valleys is used instead of the energy of the 
configuration. A similar cavity method can be developed at all temperatures,
as was done in \cite{MP_Bethe}. In general   one would expect that the
finite free energy density computed in this way be continuous
at $T=0$, which means that one should 
get back the results of the present paper by using the $T \to 0$ limit
of the finite $T$ computations.

Strictly speaking this does not happens for the model  with $J=\pm 1$. The reason is the 
following. Strictly at $T=0$ if the local fields have integer values when we start the 
iteration, they are then always  integers. 
On the contrary, starting from the same initial condition at $T\ne 0$,
after a few iterations the local fields are integer plus corrections of order $T$.  
However it is possible (and this is what happens in the RS  and in the 1RSB
 cases) that the coefficient of the corrections proportional to $T$ increases 
exponentially with the iteration number so that, after a number of iterations which is 
proportional to $\vert \ln T \vert $, one obtains a distribution which is no more concentrated near the 
integers.  

\begin{figure}[bt]
\begin{center}
\epsfig{figure=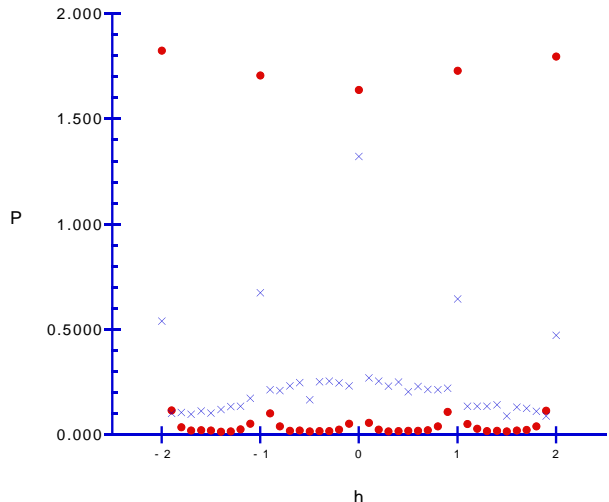,width=8cm} 
\caption{Crosses give the distribution of the cavity fields, $\cP(h)$,
computed at temperature $T=.01$ in the RS approximation.
Dots  give the site average distribution of the cavity fields, 
$\int d\cQ[P] P(h)$, at the same temperature, in the 1RSB approximation.
Clearly the distribution becomes more peaked onto integers when one 
increases the level of RSB.
\label{peak}}
\end{center}
\end{figure}

In our opinion it is quite likely that this instability is an artefact of not using the 
full RSB scheme and it should  disappear when more and more 
precise computations, with higher levels of RSB,
will be done.  In order to substantiate this point we have computed the 
distribution of the cavity fields in the $J=\pm 1$ case
for $k=2$ at very small $T$ (i.e. 
$T=0.01$) both in the RS case and in the 1RSB case. The results are 
shown in fig. \ref{peak}.
It is evident that  in the 1RSB case the distribution is much more concentrated on 
the integers.
\section*{Appendix B: Variational formulation}
It is interesting to notice that all the  equations of
stability under iteration of the cavity procedure, giving
the probability distributions of local fields, can be obtained from a variational 
principle.

In the replica symmetric case, let us define, for a generic probability
distribution $R(h)$, the functional 
\be  
\Phi[R] \equiv \Phi^{(1)}[R] -{k+1 \over 2} \Phi^{(2)}[R]  \ ,
\ee
where $ \Phi^{(1)}$ and $  \Phi^{(2)}$ are given by:
\be
\Phi^{(1)}[R]= E_{J} \int \prod_{i=1}^{k+1}\[[ dR(h_i)\]]
\min_{\sigma_0,\sigma_1,...,\sigma_{k+1}}
\(( 
-\sum_{i=1}^{k+1} J_i \sigma_0\sigma_i-\sum_{i=1}^{k+1} h_i \sigma_i
\))  
\ee
and
\be
\Phi^{(2)}[R]= E_{J} \int \
dR(h)\  dR(h')\
\min_{\sigma,,\sigma'} (-J\sigma\sigma'- h \sigma - h' \sigma' )
\ee
where $dP(h)=P(h)dh$. 
One can show, following a method similar to the one given 
in \cite{MP_Bethe}, that the
stationarity equation of this functional
\be
{\delta \Phi[R]\over \delta R(h)}=0
\ee
coincides with the equation (\ref{Piter}) giving the field distribution $P(h)$. 
Furthermore, on this solution, one can use the stationarity equation 
in order to derive:
\bea
\Phi^{(1)}[P]&=& \Delta E^{(1)}[P] \\
\Phi^{(2)}[P]&=& \Delta E^{(2)}[P]
\eea
where $\Delta E^{(1)},\Delta E^{(1)}$ are given in 
eq. (\ref{e1rs},\ref{e2rs}). This shows that, on the solution,
\be
\Phi[P]= U
\ee

The variational formulation also exists in the 1RSB case. In this case, we need to
introduce a functional $\Phi[\cQ[P(h)]]$ (a somewhat complicated object).
We introduce first the two expressions for 
 $\Phi^{(1)}_{rsb}[\cQ],\Phi^{(2)}_{rsb}[\cQ]$:
\bea
\Phi^{(1)}_{rsb}[\cQ]&=&- \frac{1}{\mu}E_{J}
 \int \prod_{i=1}^{k+1}\(( d \cQ[P_i]\))\\&&
\log \left\{  \int \prod_{i=1}^{k+1}\(( dh_i P_i(h_i)\))
\exp\[[-\mu
\min_{\sigma_0,\sigma_1,...,\sigma_{k+1}}
\((  -\sum_{i=1}^{k+1} J_i \sigma_0\sigma_i-\sum_{i=1}^{k+1} h_i \sigma_i  \)) \]] \right\} \ , 
\eea
and
\be
\Phi^{(2)}_{rsb}[\cQ]=-\frac{1}{\mu} E_{J}
\int \ d \cQ[P]\;  d \cQ[P']
\log \((  \int dh P(h) dh' P'(h')
\exp \[[-\mu\min_{\sigma,\sigma'} (-J\sigma\sigma'- h \sigma - h' \sigma' ) \]] \)) \ ,
\ee
and then construct as before:
\be  
\Phi_{rsb}[\cQ] \equiv \Phi^{(1)}_{rsb}[\cQ]
- {k+1 \over 2} \Phi^{(2)}_{rsb}[\cQ]  \ .
\ee
We must now compute the functional derivative with respect to $\cQ$.
We first notice that $\cQ$ must be a normalized distribution so that
\be
\int d\cQ [P] =1
\ee
This constraint will be enforced using a Lagrange multiplier $\lambda$.
Let us consider  the functional derivative of
$\Phi_{rsb}[\cQ]$ with 
respect to $\cQ [P]$, evaluated on an arbitrary function $P(h)=P^{*}(h)$.
We find for the first piece:
\bea
{\delta \Phi^{(1)}_{rsb}[\cQ]\over \delta \cQ [P] }&=&- \frac{k+1}{\mu}E_{J} 
 \int \prod_{i=1}^{k}\(( d \cQ[P_i]\))\\&&
\log \left\{\int dh_{k+1} P^{*}[h_{k+1}]\prod_{i=1}^{k}\(( dh_i P_i(h_i)\)) 
\exp\[[-\mu
\min_{\sigma_0,\sigma_1,...,\sigma_{k+1}}
\((  -\sum_{i=1}^{k+1} J_i \sigma_0\sigma_i-\sum_{i=1}^{k+1} h_i \sigma_i  \)) \]] \right\} \ , 
\eea
and for the second one:
\be
{\delta \Phi^{(2)}_{rsb}[\cQ]\over \delta \cQ [P] }=-\frac{2}{\mu} E_{J} 
\int \  d \cQ[P']
\log \((  \int dh P^{*}(h) dh' P'(h')
\exp \[[-\mu\min_{\sigma,\sigma'} (-J\sigma\sigma'- h \sigma - h' \sigma' ) \]] \)) \ .
\ee
Finally, the equation:
\be  
{\delta \Phi_{rsb}[\cQ]\over \delta \cQ [P] }= 
{\delta \Phi^{(1)}_{rsb}[\cQ]\over \delta \cQ [P] }
-{k+1 \over 2} {\delta \Phi^{(2)}_{rsb}[\cQ]\over \delta \cQ [P]} =   \lambda 
\ee
 should be true,  with an appropriate choice of the Lagrange 
multiplier $\lambda $, for any $P^{*}$.

In the first piece we can use:
\bea
\int \prod_{i=1}^{k}\(( dh_i P_i(h_i)\)) 
\exp\[[-\mu
\min_{\sigma_0,\sigma_1,...,\sigma_{k+1}}
\((  -\sum_{i=1}^{k+1} J_i \sigma_0\sigma_i-\sum_{i=1}^{k+1} h_i \sigma_i  \)) \]]
=\\
\int \prod_{i=1}^{k}\(( dh_i P_i(h_i)\))
\exp\[[-\mu\min_{\sigma_0,\sigma_{k+1}}\((-\sum_{i=1}^{k} a(J_i,h_i)- \sigma_0\sum_{i=1}^{k} u(J_i,h_i) -J_{k+1}  \sigma_0\sigma_{k+1}- h_{k+1}\sigma_{k+1}
  \))
 \]]
\eea
Let us then define the probability distribution $P_0(h_0)$ through:
\be
\int \prod_{i=1}^{k}\(( dh_i P_i(h_i)\))  
\exp\[[ \mu\((\sum_{i=1}^k a(J_i,h_i)\))\]] 
\delta\((h_0-\sum_{i=1}^k u(J_i,h_i)\))
=
C[\{P\}]  P_0(h_0)
\label{cdef}
\ee
where $C[\{P\}]$ is a functional of the $P_{i}$ with  $i\in\{ 1,...,k\}$.
As $\cQ[P]$ satisfies the 1RSB self-consistency equation, we know that the
 distribution $P_0(h_0)$ defined in (\ref{cdef}) is generated with a probability 
$\cQ[P_0]$. Therefore we get, on the 1RSB self-consistent-under-iteration
solution for $\cQ[P]$:
\bea
\frac {\delta \Phi^{(1)}_{rsb}[\cQ]}{\delta\cQ[P]}&=&-\frac{k+1}{\mu} E_{J} 
\int  d \cQ[P_0]
\log 
\((
\int dh_0 P_0(h_0)\; dh_{k+1}  P^*(h_{k+1})  
\right.
\\ \nonumber
&&
\left.
\exp
 \[[-\mu
   \min_{\sigma_0,\sigma_{k+1}}
   \((
      -J_{k+1}\sigma_0\sigma_{k+1}
      -h_0\sigma_0-h_{k+1}\sigma_{k+1}
   \))
 \]]
\)) \\&&
 -\frac{k+1}{\mu} E_J \int\prod_{i=1}^{k}\(( d \cQ[P_i]\)) \ln(C[\{P\}])
\eea

Clearly this is equal to $-(k+1)/2$ times  the derivative of $\Phi^{(2)}_{rsb}$, 
for any $P^*$,
up to some constant that compensates with the Lagrange multiplier $\lambda$.
The curious reader may find interesting to know that 
\be
\lambda = \frac {\delta \Phi_{rsb}[\cQ]} {\delta\cQ[P]} =
-\frac{k(k+1)}{\mu} E_J
\int \(( d\cQ[P]\)) 
\log\((  \int \(( dh\; P(h)\)) \exp\[[\mu a(J,h)
 \]]\))
\ee
 
Notice that the expression (\ref{epsfull}) for the internal energy can be
directly derived using the relation ${d (\mu \Phi(\mu)) / d \mu}=\eps $. Indeed 
\be
{d\over d \mu}\left(\mu \Phi[\cQ]\right)= \(({d\over d \mu}^{*}+{d\cQ\over d
\mu}\cdot {\delta \over \delta \cQ} \)) \left(\mu \Phi[\cQ]\right) \, 
\ee where the
derivative ${d\over d \mu}^{*}$ act only on the explicit dependence on $\mu$.
The second term in the previous equation is zero when $\cQ$ satisfies the
equilibrium condition. If we compute the derivative of eq. (\ref{final}) we get
the result contained in equation (\ref{epsfull}). It is interesting to notice
that all the equations of stability under iteration of the cavity procedure,
giving the probability distributions of local fields, can be obtained from a
variational principle. Moreover the fact that the free energy we use is a
variational one is crucial in simplifying the computation of all the
derivatives with respect to the various parameters, because only explicit
derivatives must be considered.

\section*{Appendix C: Miscellaneous comments on the definition of the states}

In this appendix we give some further comments on the definition of the LGS
and of the complexity. The subject is rather delicate from the mathematical
point of view and if one is not precise it is easy to make wrong statements
and to find contradictions. Our aim here is just to point out some of the
subtleties which one should take into account.

In the definition of LGS we have asked that the
configuration should be stable with respect to $k$ spin flips with the number
$k$ going to infinity with $N$ in some unprecised way. We could
start by defining  a LGS  of order $k$ as a configuration that is
stable with respect to $k$ spin flips. In this way one could define the
complexities $\Sigma_{k}(N)$ as $(1/N)\log \large($Number of $k$-spin-flips-stable
configurations$\large)$. The complexity of LGS could be defined as
\begin{equation}
    \Sigma_a=\lim_{k \to \infty}\lim_{N \to \infty}\Sigma_{k}(N) \ .
\end{equation}
An alternative natural definition is to define LGS as stable to changing
an infinitesimal fraction of all spins:
\begin{equation}
    \Sigma_b=\lim_{\eps \to 0}\lim_{N \to \infty}\Sigma_{\eps N}(N) \ .
\end{equation}
The question of the equivalence among various possible definitions of
 $\Sigma$, and in particular whether $\Sigma_a=\Sigma_b$, is a well posed
 mathematical problem and may have a non trivial answer.

One case in which one encounters problems when studying the large $k$, large
$N$ limit is the SK model. Although it is quite possible that these problems
are an artefact of the infinite number of neighbours in the SK model and they
may be not present on the Bethe lattice, let us mention them here briefly. In
the SK model one can compute \cite{BM,DDGGO} the complexity of one spin stable
configurations, $\lim_{N \to \infty} \Sigma_{1}(N)$. A detailed computation
shows that (at least at not too low energies) the probability distribution of
the local fields $h$ in these states $P(h)$ goes to a constant when $h$ goes
to zero. It is easy to deduce that with probability one these
states are not 2-spins stable and therefore they do not correspond to LGS,
with any reasonable definition.
This argument might seem to imply that $\lim_{N \to \infty}\Sigma_{2}(N)
<\lim_{N \to \infty}\Sigma_{1}(N)$.
However this is not true: we can count the limit for large $N$ 
of the complexity, $\hat \Sigma(\delta)$,
for the  ``$\delta$-stable configurations''
such that
\begin{equation}
    \sigma_i\sum_{k}J_{ik}\sigma_k> \delta \ .
\end{equation}
Irrespectively of the precise value of $\delta$, a simple computation
shows that $\hat \Sigma(\delta)$ is continuous at $\delta=0$, while
$\hat \Sigma(0)=\lim_{N \to \infty} \Sigma_{1}(N)$. 
On the other hand $\delta$-stable configurations are
also stable for $k$ spin flips (at least when $k\ll\delta \sqrt{N}$).
 This argument shows that for the SK model
 $\lim_{N \to \infty}\Sigma_{k}(N)$ is $k$ independent. These configurations are likely
not to be LGS if we ask for configurations that are stable
with respect to $k$ spin flips with $k \propto N^{1/2}$ but they should be 
$k$-spin-flips-stable if $k \propto N^{\gamma}$ with $\gamma<1/2$. It appears 
rather likely that there will be a change of behaviour in the complexity when
$k \sim \sqrt{N}$.
 Further and detailed investigations are needed to clarify
this point.

After this digression on the SK model, let us come back to the 
computation of the LGS complexity that we have done in this paper.
As we  have already stressed,  we do not use directly any
of the previous definitions based on $\Sigma_k(N)$.
We have just argued that a LGS must correspond to a local
solution of the cavity equations and we have counted the number of these solutions in
the infinite $N$ limit. This procedure does not give a clue on the finite $N$ 
effects, and therefore on the definition of LGS on one given
(finite $N$) sample. 

One possibility could be to write directly these equations on a finite lattice
(these are nothing but TAP equations \cite{TAP}) and to associate to each
solution of these equation a LGS. However it is well known that in general
global solutions of TAP equations for one given sample are rather rare
\cite{nemoto}: the local analysis of the solution of the equations near to a
given site of the lattice that we perform does not guarantee to provide a
global solution to TAP equations. Probably the best one can hope is the
following. We could define a quasi-solution of order $\alpha$ of TAP equations
as a configuration of spins such that a fraction $1-\alpha$ of all TAP
equations are satisfied and a fraction 
$\alpha$ of TAP equations are not
satisfied.
Our  $N \to \infty$ analysis considers on the same footing the true solutions of all
TAP equations or the quasi-solutions where a vanishing fraction (for $N \to
\infty$) of the equations is not satisfied. So one could conjecture that the
LGS correspond to quasi-solutions of order $\alpha$, in the limit where $\alpha
\to 0$. Notice that the $\alpha \to 0$ and $N \to \infty$ limits may 
not commute and therefore one should be rather careful before making any
precise statement on the number of TAP solutions.

In order to illustrate the previous point let us consider a simple example, a one
dimensional chain, which is a particular example of a Bethe lattice with $k=1$.
Let us assume that the system is antiferromagnetic (i.e $J(i,i+1)=-1$) and that
periodic boundary conditions are used.

The TAP equations are given in terms of the local cavity fields:
\begin{equation}
    h^{\leftarrow}(i)=-h^{\leftarrow}(i-1) \ \ \ ; \ \  \ 
    h^{\rightarrow}(i)=-h^{\rightarrow}(i+1) 
\end{equation}
where we have used the obvious notation
$
    h^{\leftarrow}(i)=h(i|i+1)\  ,  \ 
    h^{\rightarrow}(i)=h(i|i-1) 
$.
If $N$ is even we have two degenerate GGS, i.e.
\begin{equation}
\sigma(i)= (-1)^{i} \ \ \ \mbox{and } \ \ \ \sigma(i)= (-1)^{i+1}
\end{equation}
On the contrary if $N$ is odd we have $N$  degenerate GGS with one defect in $k$, i.e.
\begin{equation}
\sigma(i)=(-1)^{i} \ \ {\rm for} \ \ i<k \ \ ; \ \  \sigma(i)=(-1)^{i+1}\ \ {\rm for} \ \ i\ge k
\end{equation}
If we look at the TAP equations we see that for $N$ even there are two
solutions  (corresponding to each of the two GGS)
\begin{equation}
    h^{\leftarrow}(i)=
    h^{\rightarrow}(i)=\pm (-1)^{i}
\end{equation}
On the contrary there are no solutions of the TAP equations for odd $N$ as an
effect of the periodic boundary conditions. Obviously there are $N$ quasi
solutions of order $1$, that luckily correspond to  the $N$ GGS.


\end{document}